\documentclass[useAMS,usenatbib]{mn2e}
\usepackage{amsfonts}
\usepackage{bbm}
\usepackage{graphicx}
\usepackage{times}
\usepackage{graphicx}
\usepackage{rotating}
\usepackage{array}
\usepackage{epsfig}
\usepackage{textcomp}
\usepackage{amsmath}
\usepackage{natbib}

\newcommand{\Msun}{\>{\rm M_\odot}}

\newcommand{\kms}{\,{\rm km\,s^{-1}}}

\title{A Schwarzschild model of the Galactic bar with initial density from N-body simulations}

\author[Y. Wang et al.]
{Yougang Wang$^{1}$\thanks{email: wangyg@bao.ac.cn}, Shude Mao$^{2,
3}$, Richard J. Long$^{2,3}$, Juntai Shen$^{4}$\\
$^1$Key Laboratory of Optical Astronomy, National Astronomical
Observatories, Chinese Academy of Sciences, Beijing 100012, China\\
$^2$National Astronomical Observatories, Chinese Academy of
Sciences, A20 Datun Road, Chaoyang District, Beijing 100012, China\\
$^3$Jodrell Bank Centre for Astrophysics, Alan Turing Building, The
University of Manchester, Manchester M13 9PL, UK\\
$^4$Key Laboratory for Research in Galaxies and Cosmology,
Shanghai Astronomical Observatory, Chinese Academy of Sciences, \\
80 Nandan Road, Shanghai 200030, China}
\begin{document}

\date{Accepted . Received .}

\pagerange{\pageref{firstpage}--\pageref{lastpage}} \pubyear{2012}

\maketitle

\label{firstpage}
\begin{abstract}
Using the potential from N-body simulations, we construct the
Galactic bar models with the Schwarzschild method. By varying the
pattern speed and the position angle of the bar, we find that the
best-fit bar model has pattern speed $\Omega_{\rm p}=40\ \rm{km\
s^{-1}\ kpc^{-1}}$, and bar angle $\theta_{\rm bar}=45^{\circ}$.
$N$-body simulations show that the best-fit model is stable for more
than 1.5 Gyrs. Combined with the results in Wang et al. (2012), we
find that the bar angle and/or the pattern speed are not well
constrained by BRAVA data in our Schwarzschild models. The proper
motions predicted from our model are slightly larger than those
observed in four fields. In the future, more kinematic data from the
ground and space-based observations will enable us to refine our
model of the Milky Way bar.

\end{abstract}
\begin{keywords}
Galaxy: bulge -- Galaxies: kinematics and dynamics  -- galaxies:
structure
\end{keywords}

\maketitle

\section{Introduction}

Perhaps two-thirds of spiral galaxies in the Universe are barred
\citep[e.g.][]{2012ApJ...745..125L}. Although there are
several formation scenarios for the galactic bar
\citep{2013MNRAS.429.1949A,2013arXiv1304.1667S}, the issue is not
yet completely settled. Both tidal interactions and internal secular
processes have been proposed
\citep[e.g.][]{2004ARA&A..42..603K,2012arXiv1211.6752A}, although
the latter may be the dominating mechanism
\citep{2009MNRAS.397..726L}. The nearest bar is the one in our own
Galaxy. Its age and formation history still remain somewhat
uncertain, despite the large amount of photometric and kinematic
data (see below). In this context, understanding the Galactic bar
offers important clues for understanding other barred structures in
the universe.

The first evidence for a bar in our Galaxy is from 21-cm
observations \citep{1964IAUS...20..195D}. Due to extinction in
optical wavelengths, optical observation of the bar is difficult.
Direct evidence for a bar at the Galactic center comes from
near-infrared studies \citep{1991ApJ...379..631B}. Microlensing
events also favour the existence of a bar in the inner Galaxy
\citep{1994ApJ...435L.113P,1994ApJ...437L..31E,1995ApJ...440L..13Z}.
Modelling of the surface brightness in inner regions of the Milky
Way from COBE infrared observations has shown that the angle between
the major axis of the bar and the Sun-Galactic center line, $\theta_{\rm{bar}}$, is in the range of
$13.4^{\circ}-40^{\circ}$
\citep{1994PhDT.........5Z,1996MNRAS.283..149Z,1996MNRAS.282..175Z,
1997MNRAS.288..365B}. \cite{2000AJ....119..800D} used the Hercules
stream to put constraints on the bar angle $10^{\circ}<\theta_{\rm
bar}<70^{\circ}$ and the bar pattern speed $\Omega_{\rm p}=47.9\pm3.9\ \rm{km\ s^{-1}\ kpc^{-1}}$. \cite{2007ApJ...664L..31M} used the Oort
constants as a constraint, and found that $\Omega_{\rm p}=48.4\pm1.0\ \rm{km\ s^{-1}\ kpc^{-1}}$ and $20^{\circ}<\theta_{\rm
bar}<45^{\circ}$. Considering some low-velocity streams, in addition
to the Hercules stream, \cite{2010MNRAS.407.2122M} found
$\Omega_{\rm p}=47.1\pm1.8\ \rm{km\ s^{-1}\ kpc^{-1}}$ and
$30^{\circ}<\theta_{\rm bar}<45^{\circ}$. On the other hand, studies of star counts indicate that the bar angle is no more than $45^{\circ}$
\citep{1994ApJ...429L..73S,2002MNRAS.337..895M,2007MNRAS.378.1064R}. A most recent study by \cite{2013arXiv1303.6430C} indicates $\theta_{\rm
bar}=30^{\circ}$.

Confusingly, some star count observations extending to larger
longitudes found a separate, flat long bar with an orientation of
$43^{\circ}$ \citep{2005ApJ...630L.149B,2007AA...465..825C}. This is
also supported by a recent 6.7 GHz methanol maser investigation
($45^{\circ}$,  \citealt{2011arXiv1103.3913G}). A recent
simulation suggested that both the inner bar and the long thin bar found
by these studies may be manifestations of the same bar structure \citep{2011ApJ...734L..20M}.

A beneficial approach is to constrain the bar model by combining
photometric data with kinematics. Using Schwarzschild's
orbit-superposition technique, Zhao (1996, hereafter ZH96)
constructed a self-consistent bar model by using limited kinematic
data. This model reproduced the observed surface brightness,
velocity and velocity dispersion measurements in Baade's Window.
However, the radial velocity predicted from ZH96 is larger than that
observed by the Bulge Radial Velocity Assay
\citep[BRAVA,][]{2008ApJ...688.1060H}. The N-body models of
\citet{1997A&A...327..983F,1999A&A...345..787F},\cite{1999MNRAS.307..584S}
and other self-consistent models
\citep{2000MNRAS.314..433H,2004ApJ...601L.155B} also used only
limited kinematic data.

Taking into account observations of the inner regions of the Milky
Way by the Hubble Space Telescope
\citep[e.g.][]{2006MNRAS.370..435K,2008ApJ...684.1110C,2012A&A...540A..48S},
the Optical Gravitational Lensing Experiment (OGLE:
\citealt{2000AcA....50....1U,2004MNRAS.348.1439S}) and BRAVA
\citep{2007ApJ...658L..29R,2008ApJ...688.1060H,2012AJ....143...57K},
extensive higher quality kinematic bar data are now available which
can help improve bar modelling. Recently, Wang et al. (2012,hereafter
Paper I) constructed a new self-consistent bar model using the
orbit-superposition technique and utilising the BRAVA data. The
density distribution in Paper I is similar to ZH96, the main
differences being the bar angle and disk mass. The new model fits
the radial velocity and velocity dispersion well in the BRAVA
fields. However, the proper motions predicted from Paper I, along
specific Galactic longitudes, are larger than those observed. One
possible reason is that the density profile adopted in Paper I is
invalid, or there may be contaminated by disc stars. The model
density profile was obtained by modelling the surface brightness
distribution from \textit{low resolution} COBE observations.

The aim of this paper is to use an alternative density distribution
to construct a self-consistent bar model. The density distribution
adopted here is from the N-body model by
\cite{2010ApJ...720L..72S} (hereafter Shen10). Compared with the
density model in Paper I, the Shen10 model has two advantages: (1)
the model fits the BRAVA data well; (2) the model is simple. While
the density distribution in Paper I has three components, a prolate
bar, a boxy bulge and an axi-symmetric disk, the Shen10's model has
only a bulge and a spherical dark matter halo.

There are two popular methods which can be used to construct the
dynamical models. One is Schwarzschild's orbit-superposition
technique \citep{1979ApJ...232..236S}, which has been applied to
external galaxies
\citep[e.g.][]{2002A&A...388..766S,2006MNRAS.366.1126C,2008MNRAS.385..647V,2010ApJ...711..484S},
and has also been used to construct self-consistent Galactic bar
models \citep[and Paper I]{1996MNRAS.283..149Z,
2000MNRAS.314..433H}. The other is the Made-to-Measure (M2M)
algorithm
\citep{1996MNRAS.282..223S,2007spts.conf...99J,2007MNRAS.376...71D,2008MNRAS.385.1729D,
2009MNRAS.395.1079D,2010MNRAS.405..301L,2011MNRAS.415.1244D,2012MNRAS.tmp.2408L,2012MNRAS.tmp.2607M},
which was applied to the Milky Way by \cite{2004ApJ...601L.155B}.
However, only the equatorial surface brightness of the Milky Way was
used in their model: no kinematic constraints were applied and their
effective field is small. Most recently, in a companion paper, Long
et al. (2013, hereafter Long13) improved the M2M modelling, by
adapting it to a rotating frame. The density model used in Long13 is
also from Shen10's simulation. The radial velocity and velocity
dispersion predicted from the M2M model not only fit the BRAVA data,
but also match Shen10's results. Therefore, another motivation of
this paper is to compare the results from Schwarzschild's method
with those from M2M.
Throughout this paper, we adopt the velocity unit
as $\kms$, and the distance unit as kpc. Correspondingly,  the time
unit in our paper is 0.98 Gyr.

\section{model and potential}

In Shen10, the bar is evolved from an initially unbarred, thin disk.
The disk mass is $M_{\rm d}=4.25\times10^{10}\Msun$, which is
realized by $10^6$ equal mass particles. Around the disk, there is a
rigid pseudo-isothermal halo with potential $\Phi=\frac{1}{2}v_{\rm
c}^2\ln(1+\frac{{r^2}}{{R_{\rm c}^2}})$, where $V_{\rm
c}\approx 250\kms$, and $R_{\rm c}\approx 15$ kpc is the core
radius; these values are the same as those in Shen10. Regarding the
rotation curve, although in Shen10 the initial rotation curve has
$V_{\rm c}\sim 200-220\kms$ between 5 to 20 kpc, after the bar forms
the rotation curve at the solar position does change considerably to
a lower value, same as shown in the present paper (Fig.1). From the
different snapshots in simulation, Shen10 obtained a bar pattern
speed $\Omega_{\rm p}=40\ \rm{km\ s^{-1}\ kpc^{-1}}$.

\begin{figure}
\resizebox{\hsize}{!}{\includegraphics{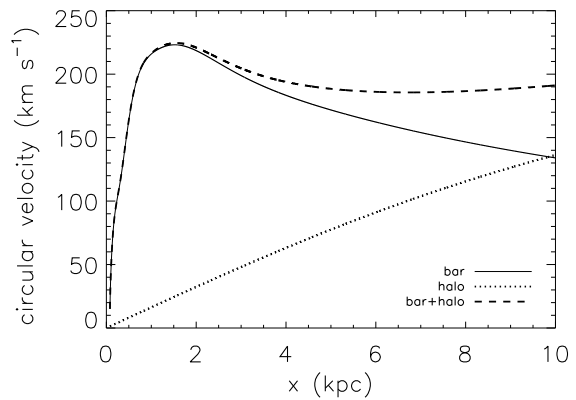}} \caption{Circular
velocity along the major axis of the bar. The solid, dotted and
dashed lines represent the bar, halo, and bar+halo, respectively. }
\label{vc} \vspace{0.5cm}
\end{figure}

In order to obtain the potential and the accelerations of the model
particles for the disk, we follow the self-consistent field method
of \cite{1992ApJ...386..375H}. The rotation curve of the model is
shown in Figure~\ref{vc} along the major axis of the bar. It can be
seen that the circular velocity is nearly constant with a value of
$v_{\rm circ}\approx190\kms$ beyond 4 kpc, which is slightly smaller
than observations in several studies
~(\citealt{1989ApJ...342..272F,2008ApJ...689.1044G,2009ApJ...700..137R,2012ApJ...759..131B},
see also \citealt{2008ApJ...679.1239W} for theoretical models), but
still consistent with those in \cite{2008ApJ...684.1143X} and
\cite{2009PASJ...61..227S}. The peak of the rotation curve in our
model is at $\sim1.5$ kpc, while that in observations is at 0.3-0.5
kpc \citep{2009PASJ...61..227S}. This difference likely originates
from the initial conditions of the N-body simulation, where the bar
is evolevd from a thin exponential disk with scale length $\sim1.9$
kpc and scale height 0.2 kpc.

%\textbf{In section 5, we will fit the galactic-centric radial
%velocity from the BRAVA survey with our models. However, the BRAVA
%survey adopts a rotation velocity of $v_{\rm circ}\approx220\kms$ in
%the solar neighbourhood \citep{2008ApJ...688.1060H}, which differs
%from $190\kms$ at 8 kpc in our simulation. Since the projection of
%the solar circular motion along the line of sight is $v_{\rm circ}
%\cos b \sin l$, the difference in the rotation velocity will cause a
%systematic offset. However, even at our largest longitude
%$|l|=10^\circ$ (and $b=0^\circ$), this will only introduce a
%systematic offset of $5.2\kms$, comparable to the error bar in our
%average velocity ($5-10\kms$), and so we do not expect this
%difference will change our results significantly.}

\section{Kinematic data}

The kinematic data used in this paper are the radial velocities and
velocity dispersions from four years of observations by the BRAVA
project. The data are described in \cite{2008ApJ...688.1060H} and
\cite{2012AJ....143...57K}. The data comprise 3 stripes along
$b=-4^{\circ}$, $-6^{\circ}$, $-8^{\circ}$ together with some fields
along the minor axis ($l=0^{\circ}$).

\section{Model construction}
\subsection{Orbit-superposition technique}
The key point of Schwarzschild's orbit superposition technique is to
construct many orbits which can be used to reproduce the input model
and other available observations. From the solutions of the
linear equations, we can obtain the weight of each orbit. Combined
with the position and velocity in orbits, we can construct the full
phase-space distribution function of the model.

The system of linear equations can be written as
\begin{equation}\label{wei1}
\mu_i=\frac{\sum_{j=1}^{\rm{N_o}}W_jO_{ij}V_{ij}}{\sum_{j=1}^{\rm{N_o}}W_jO_{ij}},~~~~~~~~
i=1,.....,\rm{N_c}
\end{equation}
where $\mu_i$ can be the volume density, surface density, the
average moment, or higher order velocity moments in each cell $i$.
$N_{\rm o}$ is the total number of orbits, $N_{\rm c}$ is the number
of the spatial cells, $O_{ij}$ is the time the orbits spends in each cell, $V_{ij}$
are velocity, and high order velocity components in each cell. Following Paper I, we divide the first
octant into 1000 cells. Due to the model symmetry, the other octants
are reflected to the first octant. In each x-, y- and z- direction,
the system is divided into 10 bins. The axis ratios of the bar in
the inner 5 kpc are 1:0.8:0.4, so each cell is a box with
$dx=0.5$\,kpc, $dy=0.4$\,kpc and $dz=0.2$ kpc.

More practically, equation~\ref{wei1} can be written as a
set of linear equations
\begin{equation}\label{wei1b}
 \sum_{j=1}^{\rm{N_{\rm o}}}(\mu_i-V_{ij})O_{ij}W_j=0,\;{\rm
 {for}}\;
 i=1,N_c
\end{equation}
We adopt the non-negative least squares (NNLS) method
\citep{1984A&A...141..171P} to solve for $W_j$ by minimising
$\chi_w^2$, where
\begin{equation}\label{wei2}
\chi_w^2=\sum_{i=1}^{\rm{N_{\rm
c}}}\bigg|\sum_{j=1}^{\rm{N_{\rm
o}}}(\mu_i-V_{ij})O_{ij}W_j\bigg|^2.
\end{equation}
The distribution of orbit weights for the solution with the
smallest $\chi_w^2$ may not be smooth, and thus artificial. Here we
use two different smoothing methods. First, we require orbits with
adjacent initial conditions to have nearly the same weight
\citep{1996ApJ...460..136M}. In this approach, Equation~\ref{wei2}
becomes

\begin{equation}\label{wei3}
\chi_w^2=\sum_{i=1}^{\rm{N_{\rm
c}}}\bigg|\sum_{j=1}^{\rm{N_{\rm
o}}}(\mu_i-V_{ij})O_{ij}W_j\bigg|^2+\lambda\sum_{j=1}^{\rm{N_{\rm
o}}}W_j^2
\end{equation}
where $\lambda$ is a positive smoothing parameter and
$\lambda=\rm{N_{\rm o}^{-2}}$.
%It is easy to prove that this
%smoothing procedure is equivalent to the system being the
%maximization of entropy $S=-\sum_{j=1}^{\rm N_o}w_j\ln[(w_j\times\rm
%{N_o})-1]$.
We vary the value of the smoothing parameter $\lambda$. However, our
results are not sensitive to reasonable variations of $\lambda$.
This is consistent with the findings in ~\cite{2010ApJ...711..484S}.
The second smoothing method we use is the same as that adopted in
ZH96. The key point of this method is that orbits with similar
integrals of motion have similar weights. We have compared the
results from these two smoothing methods and found no significant
difference between them (see also paper I). From now on, only
results from the first smoothing method will be presented.

\subsection{Initial conditions for orbits}

The bar is known to be non-spherical with `figure rotation'
\citep[e.g.][]{1997MNRAS.288..365B,1999MNRAS.304..512E,2002MNRAS.334..355D,2010arXiv1003.2489G}
. In such a system, only Jacobi's energy is an integral of motion
\citep{2008gady.book.....B}. Therefore, it is not straightforward to
construct a full phase-space density distribution. Fortunately,
three methods have been shown to be effective in generating the
initial conditions for orbits in this type of system. These methods
are described in ZH96, \cite{2000MNRAS.314..433H} and
\cite{2011ApJ...728..128D}. In Paper I, we compared the first two
methods and found that there is no significant difference between
them. Therefore, we follow Paper I and use ZH96's method to
construct the orbit library. Here we give a brief description of the
main points of this method, and refer the reader to ZH96 for more
details. The orbits are launched tangentially with a speed less than
the circular velocity from a local apgalacticon. Moreover, orbits
are launched in close pairs perpendicularly from the xz-, yz- or
xy-symmetry plane, or from the x or y axis. If two orbits are
lunched perpendicularly from one plane, the close pairs will have
opposite velocity.

We construct different models by varying the pattern speed, and for
each model we generate $\sim13000$ orbits.

%\smao{how are these different from particle conditions???}

\subsection {Orbit integration}

\begin{figure}
\resizebox{\hsize}{!}{\includegraphics{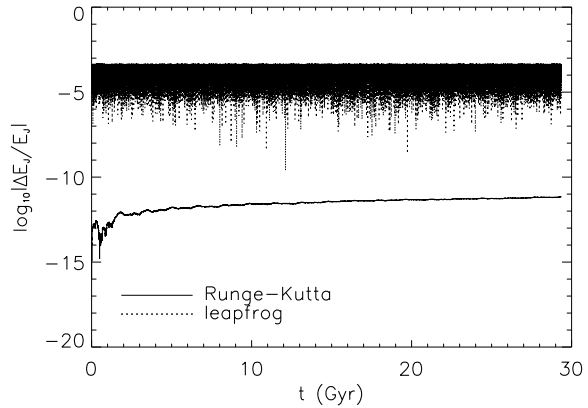}} \caption{Fractional
Jacobi's energy deviation as a function of time for 7/8 order
Runge-Kutta (solid line) and leapfrog integrator (dotted line).}
\label{de}
\end{figure}

Popular integration methods for orbits include the modified Euler,
leapfrog, Runge-Kutta and Hermite integrators. It has been shown
that the fractional energy error is the smallest for the Runge-Kutta
integrator if the integration time is short, while the leapfrog
integrator is the best method if the integration time is long
\citep{2005MNRAS.364.1105S,2008gady.book.....B}. We follow the
method proposed by \cite{2010AJ....139..803Q} to revise the general
leap-frog integrator and make it appropriate to a system with figure
rotation. We also improve the common 7/8 order Runge-Kutta
algorithm, which can give an equal time output and keeps the
Jacobi's energy as a constant with high accuracy. Figure~\ref{de}
shows the fractional Jacobi's energy deviation as a function of time
for both the 7/8 order Runge-Kutta and the leapfrog integrator. The
integration time step $\Delta t$ for the leapfrog method is
$10^{-5}$ ($\approx 10^4$ years). Clearly, the 7/8 order Runge-Kutta
integrator has higher accuracy (but with a very slow energy drift)
than the leapfrog integrator even though the integration duration is
over two Hubble times. Reducing $\Delta t$ to $10^{-6}$, the result
remains similar. Therefore, we use the 7/8 order Runge-Kutta
algorithm to carry out the orbit integrations. Every orbit is
integrated for one Hubble time, which corresponds to
$\sim$200 dynamical times in the solar neighbourhood.

\section{Results}
As shown in Paper I, the kinematics strongly depend on the pattern
speed of the bar, and the projected density strongly depends on the
bar angle. We calculate 30 models (see Table~\ref{model}) by varying
the model pattern speed and bar angle. As shown in
\cite{2002MNRAS.333..861S}, the strength of the bar is correlated
with the pattern speed, however, we have only one N-body simulation.
Therefore, we vary the pattern speed but keep the density
distribution of the bar fixed. This may be `artificial' physically.
In the context of this paper, however, it is a valid step to
experiment (keeping the density fixed). In each model, we run the
orbit-superposition technique to test the self-consistency. The
volume density, projected density, radial velocity and velocity
dispersion in four stripes $b=-4^{\circ}$, $b=-6^{\circ}$,
$b=-8^{\circ}$ and $l=0^{\circ}$ (see Figure~\ref{vr}) are used as
constraints to produce the weight of each orbit. The volume and
projected densities are derived directly from the N-body density
distribution of the bar. For volume density, only the inner 5 kpc
are considered because our aim is to check the self-consistency of
the bar. The radial velocity and velocity dispersion are from the
BRAVA survey ~\citep{2008ApJ...688.1060H,2012AJ....143...57K}. The
projected density, the radial velocity and velocity dispersion are
fitted in the range $l=[-12^{\circ},12^{\circ}]$,
$b=[-10^{\circ},10^{\circ}]$. As described in section 4.1, the orbit
weights in Equation 1 are determined by the NNLS  method and are
smoothed by following \cite{1996ApJ...460..136M} and Paper I.

\begin{figure}
\resizebox{\hsize}{!}{\includegraphics{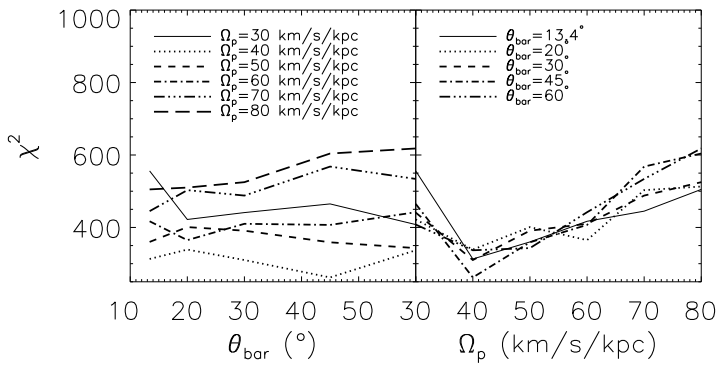}} \caption{$\chi^2$
distribution for different bar model. Left: Each line has the same pattern speed.
Right: Each line has the same bar angle.}\label{chi2}
\end{figure}

A self-consistent model should reconstruct the input volume
density and projected density well. All our models satisfy
these requirements. In order to select the best model, we calculate the
$\chi^2$ value between the observed and reconstructed radial
velocity and velocity dispersion for each model. $\chi^2$ is defined
as
\begin{equation}
\chi^2=\sum_{i=1}^{\rm{N_{\rm obs}}}\frac{(y_{\rm obs}-y_{\rm
model})^2}{\sigma_{\rm obs}^2},
\end{equation}
where $\rm{N_{\rm obs}}$ is the total number of observed data
points, $y_{\rm obs}$ and $y_{\rm model}$ are the observed and model
predicted kinematics, respectively.In Table~\ref{model}, we
show the values of $\chi^2$ for our 30 models. It is clear that the
best-fit model is Model 9. Figure~\ref{chi2} shows the results
graphically. The left panel shows the bar angle is ill constrained
while the bar pattern speed is better determined at $\Omega_{\rm
p}=40\ \rm{km\ s^{-1}\ kpc^{-1}}$. The trends are not dissimilar to
Figure 4 in \cite{2013MNRAS.428.3478L}, showing that Schwarzschild and M2M
methods are, as expected, in agreement with each other.

\begin{figure}
\resizebox{\hsize}{!}{\includegraphics{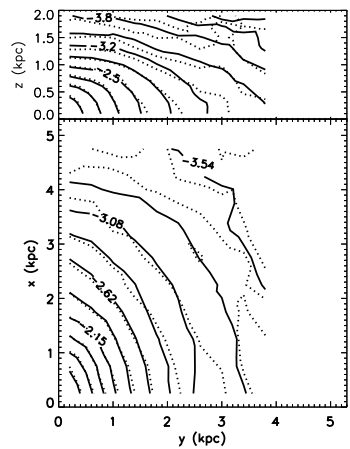}} \caption{Volume
density contours from the input model (solid lines) and from orbits
(dotted lines) in the $x-y$ (bottom panel) and $y-z$ (top panel)
planes for Model 9.} \label{iso_3d}
\end{figure}

\begin{figure}
\resizebox{\hsize}{!}{\includegraphics{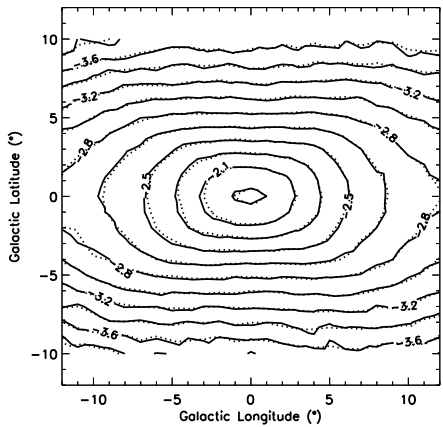}} \caption{Projected
surface density contours from the input model (solid lines) and from
orbits (dotted lines) for Model 9.} \label{iso_2d}
\end{figure}

\begin{figure}
\begin{center}
\includegraphics[height=0.35\textwidth]{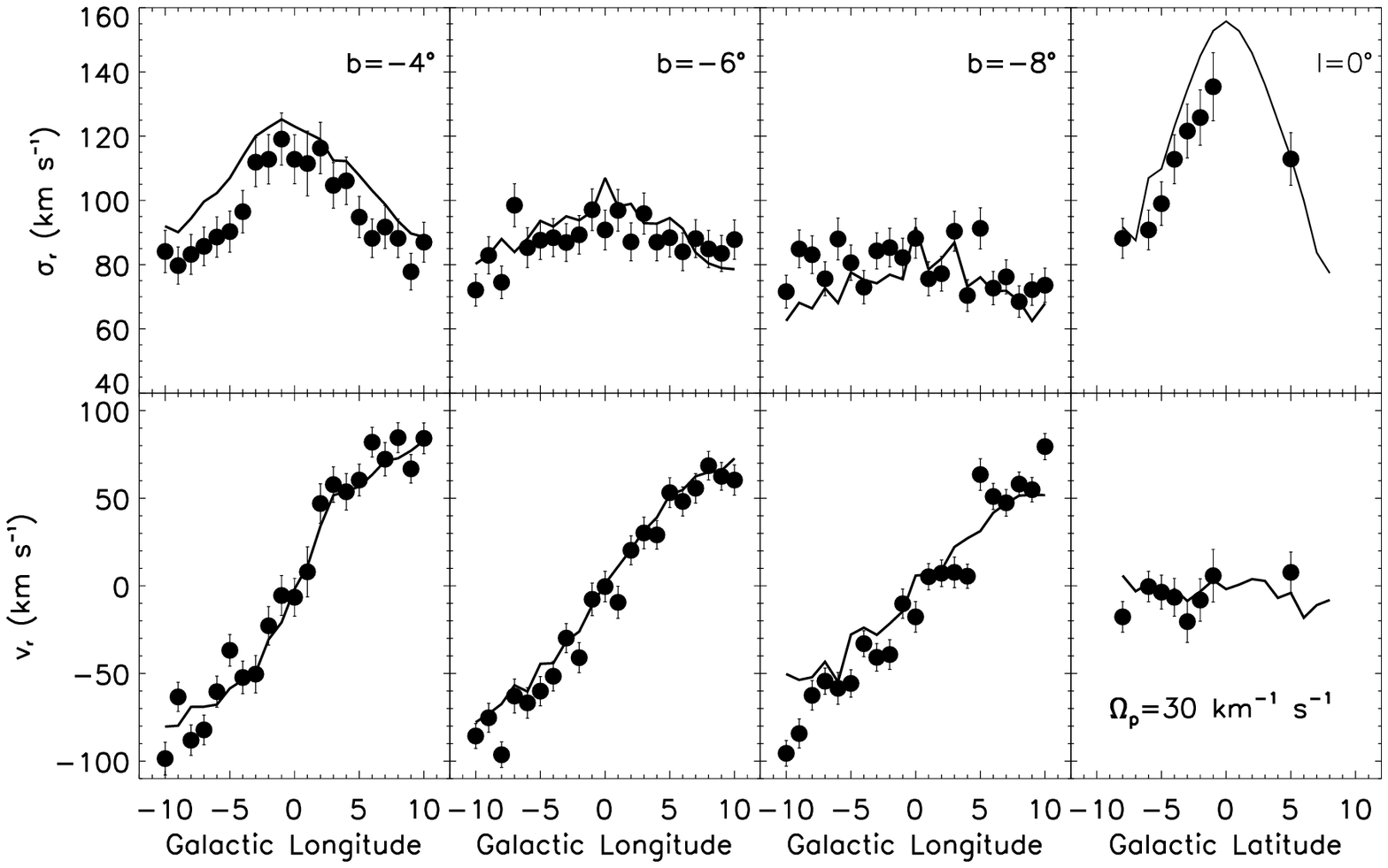}
\includegraphics[height=0.35\textwidth]{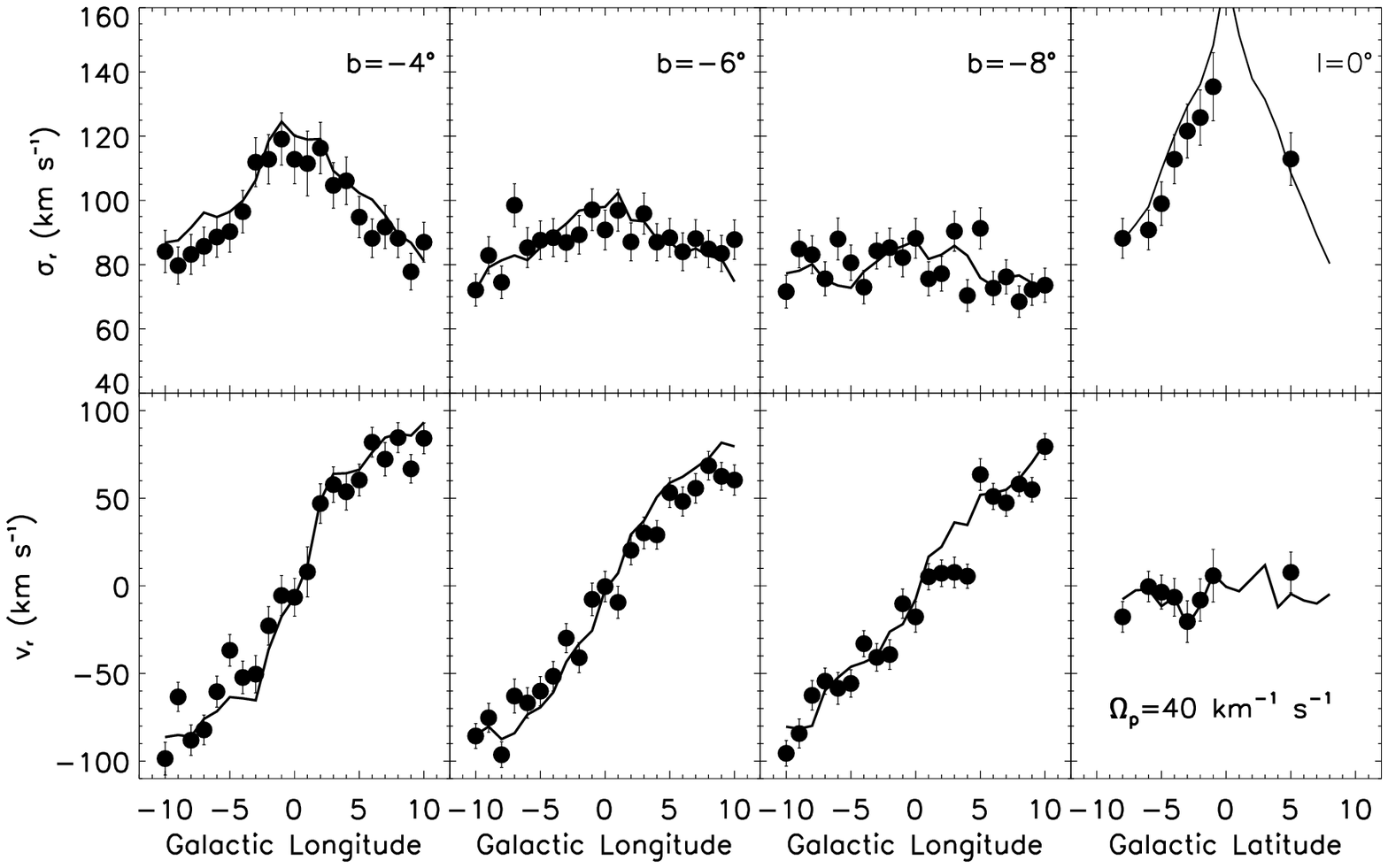}
\includegraphics[height=0.35\textwidth]{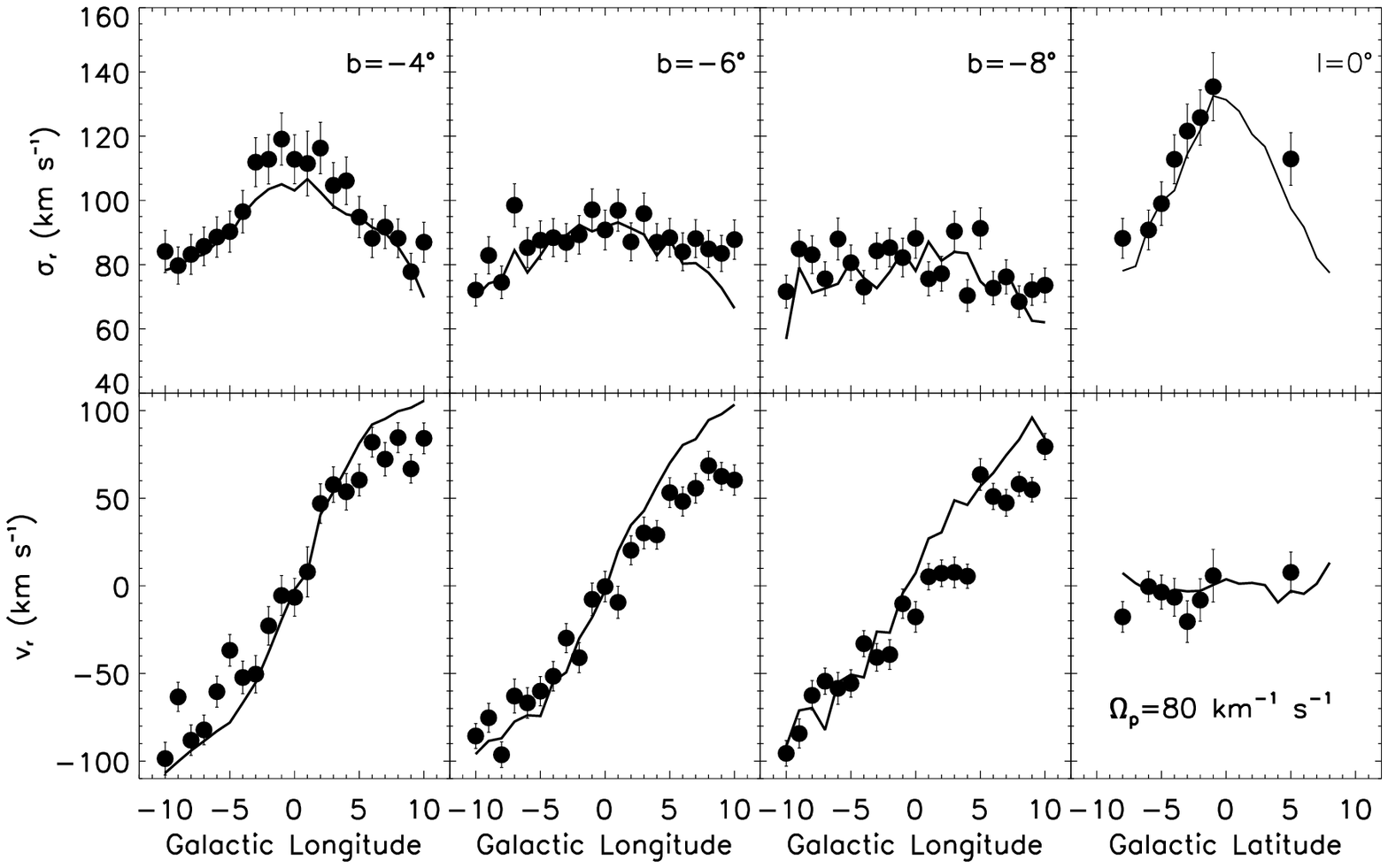}
\caption{Radial velocity and velocity dispersion distributions for
Model 4 (top panel:$\Omega_{\rm p}=30\ \rm{km\ s^{-1}\ kpc^{-1}}$,
$\theta_{\rm bar}=45^{\circ}$), Model 9 (middle panel:$\Omega_{\rm
p}=40\ \rm{km\ s^{-1}\ kpc^{-1}}$, $\theta_{\rm bar}=45^{\circ}$)
and Model 29 (bottom panel:$\Omega_{\rm p}=80\ \rm{km\ s^{-1}\
kpc^{-1}}$, $\theta_{\rm bar}=45^{\circ}$). The solid line is for
the model while the filled circles represent data from BRAVA.}
\label{vr}
\end{center}
\end{figure}

In Figures ~\ref{iso_3d} and ~\ref{iso_2d}, we show the volume
density and projected density contours, respectively, for Model 9.
In each figure, the solid and dotted lines represent the input data
and models. Clearly, Model 9 can reproduce the volume density and
projected density distributions well in most bar regions. The
deviation between the input projected density and the reconstructed
one is below $5\%$ in the inner bar region. For the volume density,
the reconstruction is not so good in the outer part. One reason is
that our input model is an N-body model, the number of particles in
the outer region is low. Notice that the surface density map
in Fig. \ref{iso_2d} does not show a pronounced asymmetry as the
corresponding surface brightness map (see Fig. 1 in Shen10) since
the latter accounts for the distance effect which amplifies the
asymmetry.

The middle panel of Figure~\ref{vr} shows the velocity and
velocity dispersion distribution for the best-fit model, Model 9. It
is seen that Model 9 can fit the BRAVA data well except a few
points. The results for  Model 4 ($\Omega_{\rm p}=30\ \rm{km\
s^{-1}\ kpc^{-1}}$, $\theta_{\rm bar}=45^{\circ}$) and Model 29
($\Omega_{\rm p}=80\ \rm{km\ s^{-1}\ kpc^{-1}}$, $\theta_{\rm
bar}=45^{\circ}$) are also given for comparison. Some wiggles in the
reconstructed kinematic profiles are clearly seen. There are two
reasons for this: (1) our input model is from an N-body simulation,
the phase space distribution is not smooth. Note that the velocity
profiles in Shen et al. (2010) are smoother than those presented
here, because they used $2^{\circ}\times 2^{\circ}$ bins while we
use $1^{\circ}\times 1^{\circ}$ bins. (2) In order to reconstruct
the phase space distribution of the Galactic bar, we did not add any
more constraints during the model fit. Therefore, regions without
data may have wiggles.

The proper motions in four fields, from observations and from the
model, for Model 9 are also given in Table~\ref{pm}. It can be seen
that the model proper motions along the longitudinal direction are
slightly larger than those observed. Comparing with the proper motions
predicted in Paper I, the proper motions predicted here are smaller,
which indicates that we have reduced the discrepancy between the
model predictions and observations. In order to make further
comparisons of the proper motions from our model with observations,
we need additional data. The predicted proper motions of Model 9 in
the range $l=[-12^{\circ},12^{\circ}]$, $b=[-10^{\circ},10^{\circ}]$
are available online \footnote{http://cosmology.bao.ac.cn/$\texttildelow$wangyg/}.

%\begin{figure}
%\resizebox{\hsize}{!}{\includegraphics{iso_3d_only_mass_fit.eps}}
%\caption{Volume density contours from the input model (solid lines)
%and from orbits (dashed lines) for Model 4. The orbits weights are
%solved by only using the three-dimensional density.}
%\label{iso_3d_only_mass}
%\end{figure}

\begin{figure}
\resizebox{\hsize}{!}{\includegraphics{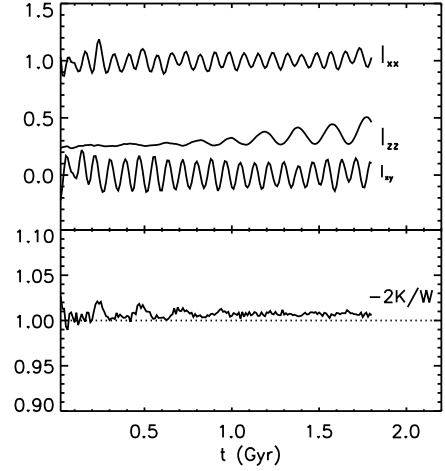}} \caption{Evolution
of $-2K/W$ (lower panel) and the three moments of inertia $I_{xx}$,
$I_{zz}$ and $I_{xy}$ for our best-fit Model 9. All the quantities
plotted here are calculated in the inertial frame, where the $x$
axis is aligned with the initial major axis of the bar.}
\label{stable}
\end{figure}

Stability is one of the important properties for an `optimal'
dynamical model. We follow the method described in Paper I to check
the stability of Model 9. Figure~\ref{stable} shows the evolution of
$-2K/W$ and three moments of inertial $I_{xx}$ (where x is defined
along the initial major axis), $I_{zz}$ (along the initial minor
axis) and $I_{xy}$ (cross term) of the bar, where $W$ is the
Claussius Virial of the system and $K$ is the total kinetic energy.
One can see that $W+2K$ satisfies the Virial theorem well. Moreover,
the cross term $I_{xy}$ in the rest frame follows nearly a
sinusoidal curve with a constant period 0.08 Gyr, which is in good
agreement with half period of the bar's rotation presented
$\pi/\Omega_{\rm p}\sim 0.08$ Gyr. Therefore, our bar model is
stable at least within 1.5 Gyrs, which is more stable than the
previous bar model in Paper I. Incidentally, this is slightly longer
than the stable period shown in other studies (e.g., Zhao 1996).

%\textbf{The velocity anisotropy parameter
%$\beta=1-\frac{1}{2}\frac{\sigma_l^2+\sigma_b^2}{\sigma_r^2}$ can
%help us to understand the orbit distribution, where $\sigma_l$ and
%$\sigma_b$ are the velocity dispersions along the Galactic longitude
%and latitude, respectively. Figure~\ref{beta} shows the velocity
%anisotropy profiles in four windows. The solid and dotted line
%represent the results of the original input model and Model 9. It
%seems that Model 9 has larger values of $\beta$ than that from the
%input N-body model, which indicates it contains more radial orbits
%than that in the input model, although the scatters are quite
%large.}

%\section{Stability of the model}
%Stability is one of the important properties for an `optimal'
%dynamical model. In order to check the stability of Model 5, we
%convert the velocities and positions of our orbits to the initial
%conditions of an N-body simulation. The total number of particles in the N-body
%simulation is $2\times10^{5}$, and they are selected by random sampling from the
%non-zero weight orbits using a process described in ZH96 and
%\cite{2009MNRAS.396..109W}. For each orbit, the number of
%particles selected is proportional to its orbit weight.

%We use the popular code GADGET-2 \citep{2005MNRAS.364.1105S} to run
%our simulation. The N-body simulation for Model 5 is run for 1.5
%Gyrs, and the position, velocity and acceleration of the particles are recorded every 0.01 Gyrs.

%In order to examine whether our model is stable, we check the
%evolution of the axis ratios and whether the virial theorem
%$W+2K=0$ (or $-2k/W=1$) is satisfied. $W$ is the Claussius Virial of the
%system and $K$ is the total kinetic energy. Figure~\ref{stab} shows
%the evolution of $-2K/W$, and the three moments of inertia $I_{xx}$,
%$I_{yy}$ and $I_{zz}$ along the three major axes of our N-body
%model. We find that the value $-2K/W$ is close to unity for the initial 0.8
%Gyrs. The axis ratios are nearly constant ($1:0.47:0.25$) over the same time period.
%Examination of the moments of inertia $I_{XX}$ and $I_{YY}$
%reveal that the bar rotates about 3 times in 0.45 Gyrs, which is in a
%good agreement with a constant pattern rotation speed $\Omega_{\rm
%p}=40\ \rm{km\ s^{-1}\ kpc^{-1}}$.

\section{Summary and Discussion}
Based on the N-body simulation in Shen10, we have constructed 30 bar
models with different pattern speeds and bar angles using
Schwarzschild's method. The $\chi^2$ values show that the best model
has $\Omega_{\rm p}=40\ \rm{km\ s^{-1}\ kpc^{-1}}$ and bar angle
$\theta_{\rm bar}=45^{\circ}$. The pattern speed is in good
agreement with those obtained from Shen's numerical simulation and
the M2M method, while the bar angle is different from but still
consistent with that obtained from the M2M method ($\theta_{\rm
bar}=30^{\circ}$, Long13), since the bar angle is weakly
constrained, as found in the N-body simulation and M2M. We
have fitted the galactic-centric radial velocity from the BRAVA
survey with our models. However, the BRAVA survey adopts a rotation
velocity of $v_{\rm circ}\approx220\kms$ in the solar neighbourhood
\citep{2008ApJ...688.1060H}, which differs from $190\kms$ at 8 kpc
in our simulation. Since the projection of the solar circular motion
along the line of sight is $v_{\rm circ} \cos b \sin l$, the
difference in the rotation velocity will cause a systematic offset.
However, even at our largest longitude $|l|=10^\circ$ (and
$b=0^\circ$), this will only introduce a systematic offset of
$5.2\kms$, comparable to the error bar in our average velocity
($5-10\kms$), and so we do not expect this difference will change
our results significantly. The best-fit model can reconstruct the
volume and projected densities of the numerical bar well, and can
also fit the observed radial velocities and velocity dispersions in
the BRAVA fields. The proper motions, along specific Galactic
longitudes, predicted by our model are slightly larger than those in
observations. Compared to the model in Paper I, our best-fit model
has better agreement of the proper motions along the Galactic
longitude with observations than Paper I. Moreover, the stability of
the bar in Model 9 is better than that in Paper I. From the studies
here and Paper I, we draw some conclusions:

\begin{enumerate}

\item The BRAVA data can be fitted by models within large
parameter ranges - the pattern speed can be from $\Omega_{\rm p}=40\
\rm{km\ s^{-1}\ kpc^{-1}}$ to $\Omega_{\rm p}=80\ \rm{km\ s^{-1}\
kpc^{-1}}$, and the bar angle from $\theta_{\rm bar}=13.4^{\circ}$
to $\theta_{\rm bar}=60^{\circ}$. In other words, it is not easy to
say which model is the best-fit model from the BRAVA data alone.
Moreover, it seems that there is a degeneracy between the pattern
speed and the density distribution of the model. In paper I,
$\Omega_{\rm p}=60\ \rm{km\ s^{-1}\ kpc^{-1}}$ can fit the data
well, while $\Omega_{\rm p}=40\ \rm{km\ s^{-1}\ kpc^{-1}}$ can best
fit the data here.

%\item \textbf{The bar model is only stable for 0.8 Gyr for our best-fit Model.
%One possible reason is that we only consider the self-consistency of
%the bar region, while ideally the whole region should be modelled.
%Another possible reason is that more than $90\%$ of the orbits are
%irregular in this model. The fraction of irregular orbits depends on
%the potential. Therefore, we need to use other density models to
%construct a self-consistent and stable Galactic bar model.}

\item Observed proper motions can help us to reduce the model
degeneracy. A comparison indicates that the best fit with the
observed data is $\Omega_{\rm p}=40\ \rm{km\ s^{-1}\ kpc^{-1}}$.
However, the proper motions along Galactic longitudes predicted by
our model are slightly larger than those in observations. One possible reason
is that the density model adopted is not yet sufficiently
representative of the Milky Way. At present, most density models of
the bar are constructed by using only the surface brightness, star
counts, or from N-body simulations. We need to construct the
density distribution of the bar model by combining all of them,
especially if the proper motion constraints are to be matched.
Another possible reason is that the proper motion data in the bar
may be contaminated by disk stars. We also note that proper
  motions of the same field from different observations give somewhat
  discrepant values (e.g., see Table 2 for values for Baade's window).
\end{enumerate}

Clearly more proper motion data are desirable. It will also be
important to consider other probes of kinematics, such as those
considered by Dehnen (2000, for more recent data, see Liu et al. 2012)
to constrain bar
parameters. Future data from APOGEE \citep{2011AJ....142...72E} and
ARGOS \citep{2013MNRAS.428.3660F} will provide new chemo-dynamical
constraints on the Galactic bar, which will challenge all
current modelling methods and provide strong clues about how did the bar in our
Galaxy form.

%\item The bar angle in this paper is $\theta_{\rm bar}=60^{\circ}$,
%which is larger than is determined elsewhere in other papers. The bar is evolved from
%a disk in Shen10.  As a consequence, the bar in this paper may not match that which is
%observed. However, it can still be used to study the
%structure of possible real bars, and help us gain understanding of
%bar formation and evolution processes.

%\citep{2012ApJ...753L..24L,2013MNRAS.428.3478L,2012MNRAS.427.1429W,1996MNRAS.283..149Z}
\clearpage

\begin{table}
\caption{$\chi^2$ from fitting the velocity and velocity dispersion
of BRAVA data for different input models.}

\label{model}
\begin{center}
\begin{tabular}{cccccccccccccc}\hline
 Model ID &$\Omega_{\rm p}$  &$\theta_{\rm bar}$&${\chi^2}$\\
          &$(\rm{km\ s^{-1}\ kpc^{-1}})$ &$(^\circ)$& \\

 \hline\hline
 1 & 30    &13.4  &556    \\
 2 & 30    &20    &422    \\
 3 & 30    &30    &441    \\
 4 & 30    &45    &465    \\
 5 & 30    &60    &408    \\
  \hline
 6 & 40    &13.4  &313    \\
 7 & 40    &20    &339    \\
 8 & 40    &30    &310   \\
 9 & 40    &45    &262   \\
 10 & 40   &60    &337    \\
  \hline
 11 & 50    &13.4  &360    \\
 12 & 50    &20    &401    \\
 13 & 50    &30    &391    \\
 14 & 50    &45    &359    \\
 15 & 50    &60    &343    \\
  \hline
 16& 60    &13.4  &417   \\
 17& 60    &20    &365    \\
 18& 60    &30    &410    \\
 19& 60    &45    &407    \\
 20& 60    &60    &442    \\
  \hline
 21&70     &13.4  &445   \\
 22&70     &20    &503    \\
 23&70     &30    &488    \\
 24&70     &45    &568    \\
 25&70     &60    &534    \\
  \hline
 26&80     &13.4  &505    \\
 27&80     &20    &510  \\
 28&80     &30    &525    \\
 29&80     &45    &604    \\
 30&80     &60    &618    \\
\hline
% 26& Shen10&13.4 &-- &--&307 & 133\\
% 27& Shen10&20   &-- &--&325 & 104\\
% 28& Shen10&30   &-- &--&287 & 133\\
% 29& Shen10&45   &-- &--&304 & 544\\
% 30& Shen10&60   &-- &--&309 & 197\\
% 31& Shen10&70   &370 & 249\\

 \hline
 \end{tabular}
\end{center}
%{\footnotesize
% \noindent
% $^{a}$()Fitting the proper motion in (1-4),(0,-8),(1,-3),(0,-6). \\

%}

\end{table}
\clearpage

\begin{table}
\caption{Observed and predicted proper motion dispersions in some
fields, the bottom four rows are predictions from Paper
I.}\label{pm}
\begin{center}
\begin{tabular}{lllllllll}\hline
 Field & (l,b)&$\sigma_l$& $\sigma_b$ & Ref. \\ \hline\hline
       & $(^{\circ})$&$(\rm{mas\ yr^{-1}})$  &$(\rm{mas\ yr^{-1}})$&  \\ \hline
 Baade's Window & (1,-4)  &$3.2\pm0.1$  &$2.8\pm0.1$ & \cite{1992AJ....103..297S}\\
 Baade's Window & (1,-4)  &$3.14\pm0.11$  &$2.74\pm0.08$ & \cite{1996ApJ...470..506Z}\\
 Baade's Window & (1.13,-3.77)  &2.9  &2.5 & \cite{2002AJ....124.2054K}\\
 Baade's Window & (1,-4)  &$2.87\pm0.08$  &$2.59\pm0.08$ & \cite{2006MNRAS.370..435K}\\
 Baade's Window & (0.9,-4)  &$3.06\pm0.11$  &$2.79\pm0.13$ &\cite{2007ApJ...665L..31S}\\
 Baade's Window & (1,-4)  &$3.13\pm0.16$  &$2.50\pm0.10$ &{\cite{2010A&A...519A..77B}}\\
 Baade's Window & (1.13,-3.76)  &$3.08\pm 0.08 $  &$2.74\pm 0.13$ &{\cite{2012A&A...540A..48S}}\\
 Plaut's Window & (0,-8)  &$3.39\pm 0.11$  &$2.91\pm 0.09$ & \cite{2007AJ....134.1432V,2009RMxAC..35..123V}  \\
 Sagittarius I & (1.25,-2.65)  &3.3  &2.7 & \cite{2002AJ....124.2054K}\\
 Sagittarius I & (1.27,-2.66)  &$3.07\pm 0.08$ &$2.73\pm 0.07$& \cite{2006MNRAS.370..435K}\\
 Sagittarius I &(1.25,-2.65)&3.067 &2.760 & \cite{2008ApJ...684.1110C}\\
 Sagittarius I & (1.26,-2.65)  &$3.11\pm 0.08 $  &$2.71\pm 0.08$ &{\cite{2012A&A...540A..48S}}  \\
 NGC 6558 & (0.28,-6.17)  &$2.45\pm 0.11$ &$2.37\pm 0.13$ & {\cite{2012A&A...540A..48S}}\\
 %\hline
 %Baade's Window & (1,-4)         &3.28  &2.38 & Model 28\\
 %Plaut's Window & (0,-8)         &3.35  &2.11 & Model 28\\
 %Sagittarius I & (1,-3)          &3.33  &2.47 & Model 28  \\
 %NGC 6558 &(0,-6)                &3.26  &2.17 & Model 28\\
 \hline
 Baade's Window & (1,-4)         &3.74  &2.49 & Model 9\\
 Plaut's Window & (0,-8)         &3.43  &2.40 & Model 9\\
 Sagittarius I & (1,-3)          &3.61  &2.58 & Model 9\\
 NGC 6558 &(0,-6)                &3.67  &2.48 & Model 9\\
 \hline
 Baade's Window & (1,-4)         &4.44  &2.52 & Wang et al. (2012)\\
 Plaut's Window & (0,-8)         &5.28  &2.32 & Wang et al. (2012)\\
 Sagittarius I & (1,-3)          &4.43  &2.67 & Wang et al. (2012)\\
 NGC 6558 &(0,-6)                &4.46  &2.36 & Wang et al. (2012)\\

 \hline
 \end{tabular}
\end{center}
% {\footnotesize
% \noindent
% $^{*}$ Proper motion data we used to compare with models
% predictions.}
% Equinox 2000.0 \\
% $^{b}$ Exposure time after lightcurve screening (see $\S$4.2) \\
% $^{c}$ Detector on the aim point \\}
\end{table}
\clearpage

%\begin{figure}
%\resizebox{\hsize}{!}{\includegraphics{fig8.ps}} \caption{Velocity
%anisotropy distributions in four windows. The solid and dotted lines
%represent the results of the original input N-body model and Model 9
%in this paper, respectively.} \label{beta}
%\end{figure}
\section*{Acknowledgments}
We thank the referee for comments and suggestions that improved the
paper. We acknowledge helpful discussions with Simon White. We also
thank Jie Wang and Bin Yue, who helped us to revise the Gadget code
and run N-body simulations. This work was started during the 2011
workshop on the Galactic bulge and bar in the Aspen Center for
Physics. We thank the hospitality of the Aspen Center for Physics,
which is supported by the NSF grant 1066293. YGW acknowledges the
support by the National Science Foundation of China (Grant No.
Y011061001 and No. Y122071001). SM and RJL acknowledge the financial
support of the Chinese Academy of Sciences and NAOC. JS acknowledges
the support by the National Science Foundation of China (Grant no.
11073037) and by the 973 Program of China (Grant no. 2009CB824800).
Computer runs were performed on the \emph{Laohu } computer cluster
of NAOC and the Supercomputing Center of Chinese Academy of
Sciences.

\bibliographystyle{mn2e}

\bibliography{bar}

\begin{thebibliography}{}

\bibitem[\protect\citeauthoryear{{Athanassoula}}{{Athanassoula}}{2012}]{2012ar%
Xiv1211.6752A}
{Athanassoula} E.,  2012, ArXiv e-prints

\bibitem[\protect\citeauthoryear{{Athanassoula}, {Machado} \&
  {Rodionov}}{{Athanassoula} et~al.}{2013}]{2013MNRAS.429.1949A}
{Athanassoula} E.,  {Machado} R.~E.~G.,    {Rodionov} S.~A.,  2013, \mnras,
  429, 1949

\bibitem[\protect\citeauthoryear{{Babusiaux}, {G{\'o}mez}, {Hill}, {Royer},
  {Zoccali}, {Arenou}, {Fux}, {Lecureur}, {Schultheis}, {Barbuy}, {Minniti} \&
  {Ortolani}}{{Babusiaux} et~al.}{2010}]{2010A&A...519A..77B}
{Babusiaux} C.,  {G{\'o}mez} A.,  {Hill} V.,  {Royer} F.,  {Zoccali} M.,
  {Arenou} F.,  {Fux} R.,  {Lecureur} A.,  {Schultheis} M.,  {Barbuy} B.,
  {Minniti} D.,    {Ortolani} S.,  2010, \aap, 519, A77

\bibitem[\protect\citeauthoryear{{Benjamin}, {Churchwell}, {Babler} \&
  {Indebetouw}}{{Benjamin} et~al.}{2005}]{2005ApJ...630L.149B}
{Benjamin} R.~A.,  {Churchwell} E.,  {Babler} B.~L.,    {Indebetouw} e.~a.,
  2005, \apjl, 630, L149

\bibitem[\protect\citeauthoryear{{Binney}, {Gerhard} \& {Spergel}}{{Binney}
  et~al.}{1997}]{1997MNRAS.288..365B}
{Binney} J.,  {Gerhard} O.,    {Spergel} D.,  1997, \mnras, 288, 365

\bibitem[\protect\citeauthoryear{{Binney} \& {Tremaine}}{{Binney} \&
  {Tremaine}}{2008}]{2008gady.book.....B}
{Binney} J.,  {Tremaine} S.,  2008, {Galactic Dynamics: Second Edition}.
Princeton University Press

\bibitem[\protect\citeauthoryear{{Bissantz}, {Debattista} \&
  {Gerhard}}{{Bissantz} et~al.}{2004}]{2004ApJ...601L.155B}
{Bissantz} N.,  {Debattista} V.~P.,    {Gerhard} O.,  2004, \apjl, 601, L155

\bibitem[\protect\citeauthoryear{{Blitz} \& {Spergel}}{{Blitz} \&
  {Spergel}}{1991}]{1991ApJ...379..631B}
{Blitz} L.,  {Spergel} D.~N.,  1991, \apj, 379, 631

\bibitem[\protect\citeauthoryear{{Bovy} et~al.}{2012}]{2012ApJ...759..131B}
{Bovy} J.,  {Allende Prieto} C.,  {Beers} T.~C.,  {Bizyaev} D.,  {da Costa}
  L.~N.,  {Cunha} K.,  {Ebelke} G.~L.,  {Eisenstein} D.~J.,  {Frinchaboy}
  P.~M.,  {Garc{\'{\i}}a P{\'e}rez} A.~E.,  {Girardi} L.,  {Hearty} F.~R.,
  {Hogg} D.~W.,  {Holtzman} J.,  {Maia} M.~A.~G.,  {Majewski} S.~R.,
  {Malanushenko} E.,  {Malanushenko} V.,  {M{\'e}sz{\'a}ros} S.,  {Nidever}
  D.~L.,  {O'Connell} R.~W.,  {O'Donnell} C.,  {Oravetz} A.,  {Pan} K.,
  {Rocha-Pinto} H.~J.,  {Schiavon} R.~P.,  {Schneider} D.~P.,  {Schultheis} M.,
   {Skrutskie} M.,  {Smith} V.~V.,  {Weinberg} D.~H.,  {Wilson} J.~C.,
  {Zasowski} G.,  2012, \apj, 759, 131

\bibitem[\protect\citeauthoryear{{Cabrera-Lavers}, {Hammersley},
  {Gonz{\'a}lez-Fern{\'a}ndez}, {L{\'o}pez-Corredoira}, {Garz{\'o}n} \&
  {Mahoney}}{{Cabrera-Lavers} et~al.}{2007}]{2007AA...465..825C}
{Cabrera-Lavers} A.,  {Hammersley} P.~L.,  {Gonz{\'a}lez-Fern{\'a}ndez} C.,
  {L{\'o}pez-Corredoira} M.,  {Garz{\'o}n} F.,    {Mahoney} T.~J.,  2007, \aap,
  465, 825

\bibitem[\protect\citeauthoryear{{Cao}, {Mao}, {Nataf}, {Rattenbury} \&
  {Gould}}{{Cao} et~al.}{2013}]{2013arXiv1303.6430C}
{Cao} L.,  {Mao} S.,  {Nataf} D.,  {Rattenbury} N.~J.,    {Gould} A.,  2013,
  ArXiv e-prints

\bibitem[\protect\citeauthoryear{{Cappellari}, {Bacon}, {Bureau}, {Damen},
  {Davies}, {de Zeeuw}, {Emsellem}, {Falc{\'o}n-Barroso}, {Krajnovi{\'c}},
  {Kuntschner}, {McDermid}, {Peletier}, {Sarzi}, {van den Bosch} \& {van de
  Ven}}{{Cappellari} et~al.}{2006}]{2006MNRAS.366.1126C}
{Cappellari} M.,  {Bacon} R.,  {Bureau} M.,  {Damen} M.~C.,  {Davies} R.~L.,
  {de Zeeuw} P.~T.,  {Emsellem} E.,  {Falc{\'o}n-Barroso} J.,  {Krajnovi{\'c}}
  D.,  {Kuntschner} H.,  {McDermid} R.~M.,  {Peletier} R.~F.,  {Sarzi} M.,
  {van den Bosch} R.~C.~E.,    {van de Ven} G.,  2006, \mnras, 366, 1126

\bibitem[\protect\citeauthoryear{{Clarkson}, {Sahu}, {Anderson}, {Smith},
  {Brown}, {Rich}, {Casertano}, {Bond}, {Livio}, {Minniti}, {Panagia},
  {Renzini}, {Valenti} \& {Zoccali}}{{Clarkson}
  et~al.}{2008}]{2008ApJ...684.1110C}
{Clarkson} W.,  {Sahu} K.,  {Anderson} J.,  {Smith} T.~E.,  {Brown} T.~M.,
  {Rich} R.~M.,  {Casertano} S.,  {Bond} H.~E.,  {Livio} M.,  {Minniti} D.,
  {Panagia} N.,  {Renzini} A.,  {Valenti} J.,    {Zoccali} M.,  2008, \apj,
  684, 1110

\bibitem[\protect\citeauthoryear{{Das}, {Gerhard}, {Mendez}, {Teodorescu} \&
  {de Lorenzi}}{{Das} et~al.}{2011}]{2011MNRAS.415.1244D}
{Das} P.,  {Gerhard} O.,  {Mendez} R.~H.,  {Teodorescu} A.~M.,    {de Lorenzi}
  F.,  2011, \mnras, 415, 1244

\bibitem[\protect\citeauthoryear{{de Lorenzi}, {Debattista}, {Gerhard} \&
  {Sambhus}}{{de Lorenzi} et~al.}{2007}]{2007MNRAS.376...71D}
{de Lorenzi} F.,  {Debattista} V.~P.,  {Gerhard} O.,    {Sambhus} N.,  2007,
  \mnras, 376, 71

\bibitem[\protect\citeauthoryear{{de Lorenzi}, {Gerhard}, {Saglia}, {Sambhus},
  {Debattista}, {Pannella} \& {M{\'e}ndez}}{{de Lorenzi}
  et~al.}{2008}]{2008MNRAS.385.1729D}
{de Lorenzi} F.,  {Gerhard} O.,  {Saglia} R.~P.,  {Sambhus} N.,  {Debattista}
  V.~P.,  {Pannella} M.,    {M{\'e}ndez} R.~H.,  2008, \mnras, 385, 1729

\bibitem[\protect\citeauthoryear{{de Vaucouleurs}}{{de
  Vaucouleurs}}{1964}]{1964IAUS...20..195D}
{de Vaucouleurs} G.,  1964, in {F.~J.~Kerr} ed., The Galaxy and the Magellanic
  Clouds Vol.~20 of IAU Symposium, {Interpretation of velocity distribution of
  the inner regions of the Galaxy}.
p.~195

\bibitem[\protect\citeauthoryear{{Debattista}, {Gerhard} \&
  {Sevenster}}{{Debattista} et~al.}{2002}]{2002MNRAS.334..355D}
{Debattista} V.~P.,  {Gerhard} O.,    {Sevenster} M.~N.,  2002, \mnras, 334,
  355

\bibitem[\protect\citeauthoryear{{Dehnen}}{{Dehnen}}{2000}]{2000AJ....119..800%
D}
{Dehnen} W.,  2000, \aj, 119, 800

\bibitem[\protect\citeauthoryear{{Dehnen}}{{Dehnen}}{2009}]{2009MNRAS.395.1079%
D}
{Dehnen} W.,  2009, \mnras, 395, 1079

\bibitem[\protect\citeauthoryear{{Deibel}, {Valluri} \& {Merritt}}{{Deibel}
  et~al.}{2011}]{2011ApJ...728..128D}
{Deibel} A.~T.,  {Valluri} M.,    {Merritt} D.,  2011, \apj, 728, 128

\bibitem[\protect\citeauthoryear{{Eisenstein}, {Weinberg}, {Agol}, {Aihara},
  {Allende Prieto}, {Anderson}, {Arns}, {Aubourg}, {Bailey}, {Balbinot} \& et
  al.}{{Eisenstein} et~al.}{2011}]{2011AJ....142...72E}
{Eisenstein} D.~J.,  {Weinberg} D.~H.,  {Agol} E.,  {Aihara} H.,  {Allende
  Prieto} C.,  {Anderson} S.~F.,  {Arns} J.~A.,  {Aubourg} {\'E}.,  {Bailey}
  S.,  {Balbinot} E.,    et al. 2011, \aj, 142, 72

\bibitem[\protect\citeauthoryear{{Englmaier} \& {Gerhard}}{{Englmaier} \&
  {Gerhard}}{1999}]{1999MNRAS.304..512E}
{Englmaier} P.,  {Gerhard} O.,  1999, \mnras, 304, 512

\bibitem[\protect\citeauthoryear{{Evans}}{{Evans}}{1994}]{1994ApJ...437L..31E}
{Evans} N.~W.,  1994, \apjl, 437, L31

\bibitem[\protect\citeauthoryear{{Fich}, {Blitz} \& {Stark}}{{Fich}
  et~al.}{1989}]{1989ApJ...342..272F}
{Fich} M.,  {Blitz} L.,    {Stark} A.~A.,  1989, \apj, 342, 272

\bibitem[\protect\citeauthoryear{{Freeman}, {Ness}, {Wylie-de-Boer},
  {Athanassoula}, {Bland-Hawthorn}, {Asplund}, {Lewis}, {Yong}, {Lane}, {Kiss}
  \& {Ibata}}{{Freeman} et~al.}{2013}]{2013MNRAS.428.3660F}
{Freeman} K.,  {Ness} M.,  {Wylie-de-Boer} E.,  {Athanassoula} E.,
  {Bland-Hawthorn} J.,  {Asplund} M.,  {Lewis} G.,  {Yong} D.,  {Lane} R.,
  {Kiss} L.,    {Ibata} R.,  2013, \mnras, 428, 3660

\bibitem[\protect\citeauthoryear{{Fux}}{{Fux}}{1997}]{1997A&A...327..983F}
{Fux} R.,  1997, \aap, 327, 983

\bibitem[\protect\citeauthoryear{{Fux}}{{Fux}}{1999}]{1999A&A...345..787F}
{Fux} R.,  1999, \aap, 345, 787

\bibitem[\protect\citeauthoryear{{Gerhard}}{{Gerhard}}{2010}]{2010arXiv1003.24%
89G}
{Gerhard} O.,  2010, ArXiv e-prints

\bibitem[\protect\citeauthoryear{{Ghez}, {Salim}, {Weinberg}, {Lu}, {Do},
  {Dunn}, {Matthews}, {Morris}, {Yelda}, {Becklin}, {Kremenek}, {Milosavljevic}
  \& {Naiman}}{{Ghez} et~al.}{2008}]{2008ApJ...689.1044G}
{Ghez} A.~M.,  {Salim} S.,  {Weinberg} N.~N.,  {Lu} J.~R.,  {Do} T.,  {Dunn}
  J.~K.,  {Matthews} K.,  {Morris} M.~R.,  {Yelda} S.,  {Becklin} E.~E.,
  {Kremenek} T.,  {Milosavljevic} M.,    {Naiman} J.,  2008, \apj, 689, 1044

\bibitem[\protect\citeauthoryear{{Green}, {Caswell}, {McClure-Griffiths},
  {Avison}, {Breen}, {Burton}, {Ellingsen}, {Fuller}, {Gray}, {Pestalozzi},
  {Thompson} \& {Voronkov}}{{Green} et~al.}{2011}]{2011arXiv1103.3913G}
{Green} J.~A.,  {Caswell} J.~L.,  {McClure-Griffiths} N.~M.,  {Avison} A.,
  {Breen} S.~L.,  {Burton} M.~G.,  {Ellingsen} S.~P.,  {Fuller} G.~A.,  {Gray}
  M.~D.,  {Pestalozzi} M.,  {Thompson} M.~A.,    {Voronkov} M.~A.,  2011, ArXiv
  e-prints

\bibitem[\protect\citeauthoryear{{H{\"a}fner}, {Evans}, {Dehnen} \&
  {Binney}}{{H{\"a}fner} et~al.}{2000}]{2000MNRAS.314..433H}
{H{\"a}fner} R.,  {Evans} N.~W.,  {Dehnen} W.,    {Binney} J.,  2000, \mnras,
  314, 433

\bibitem[\protect\citeauthoryear{{Hernquist} \& {Ostriker}}{{Hernquist} \&
  {Ostriker}}{1992}]{1992ApJ...386..375H}
{Hernquist} L.,  {Ostriker} J.~P.,  1992, \apj, 386, 375

\bibitem[\protect\citeauthoryear{{Howard}, {Rich}, {Reitzel}, {Koch}, {De
  Propris} \& {Zhao}}{{Howard} et~al.}{2008}]{2008ApJ...688.1060H}
{Howard} C.~D.,  {Rich} R.~M.,  {Reitzel} D.~B.,  {Koch} A.,  {De Propris} R.,
    {Zhao} H.,  2008, \apj, 688, 1060

\bibitem[\protect\citeauthoryear{{Jourdeuil} \& {Emsellem}}{{Jourdeuil} \&
  {Emsellem}}{2007}]{2007spts.conf...99J}
{Jourdeuil} E.,  {Emsellem} E.,  2007, in {M.~Kissler-Patig, J.~R.~Walsh, \&
  M.~M.~Roth} ed., Science Perspectives for 3D Spectroscopy {Scalable N-body
  Code for the Modeling of Early-type Galaxies}.
p.~99

\bibitem[\protect\citeauthoryear{{Kormendy} \& {Kennicutt} Jr.}{{Kormendy} \&
  {Kennicutt}}{2004}]{2004ARA&A..42..603K}
{Kormendy} J.,  {Kennicutt} Jr. R.~C.,  2004, \araa, 42, 603

\bibitem[\protect\citeauthoryear{{Koz{\l}owski}, {Wo{\'z}niak}, {Mao}, {Smith},
  {Sumi}, {Vestrand} \& {Wyrzykowski}}{{Koz{\l}owski}
  et~al.}{2006}]{2006MNRAS.370..435K}
{Koz{\l}owski} S.,  {Wo{\'z}niak} P.~R.,  {Mao} S.,  {Smith} M.~C.,  {Sumi} T.,
   {Vestrand} W.~T.,    {Wyrzykowski} {\L}.,  2006, \mnras, 370, 435

\bibitem[\protect\citeauthoryear{{Kuijken} \& {Rich}}{{Kuijken} \&
  {Rich}}{2002}]{2002AJ....124.2054K}
{Kuijken} K.,  {Rich} R.~M.,  2002, \aj, 124, 2054

\bibitem[\protect\citeauthoryear{{Kunder}, {Koch}, {Rich}, {de Propris},
  {Howard}, {Stubbs}, {Johnson}, {Shen}, {Wang}, {Robin}, {Kormendy}, {Soto},
  {Frinchaboy}, {Reitzel}, {Zhao} \& {Origlia}}{{Kunder}
  et~al.}{2012}]{2012AJ....143...57K}
{Kunder} A.,  {Koch} A.,  {Rich} R.~M.,  {de Propris} R.,  {Howard} C.~D.,
  {Stubbs} S.~A.,  {Johnson} C.~I.,  {Shen} J.,  {Wang} Y.,  {Robin} A.~C.,
  {Kormendy} J.,  {Soto} M.,  {Frinchaboy} P.,  {Reitzel} D.~B.,  {Zhao} H.,
  {Origlia} L.,  2012, \aj, 143, 57

\bibitem[\protect\citeauthoryear{{Lee}, {Park}, {Lee} \& {Choi}}{{Lee}
  et~al.}{2012}]{2012ApJ...745..125L}
{Lee} G.-H.,  {Park} C.,  {Lee} M.~G.,    {Choi} Y.-Y.,  2012, \apj, 745, 125

\bibitem[\protect\citeauthoryear{{Li}, {Gadotti}, {Mao} \& {Kauffmann}}{{Li}
  et~al.}{2009}]{2009MNRAS.397..726L}
{Li} C.,  {Gadotti} D.~A.,  {Mao} S.,    {Kauffmann} G.,  2009, \mnras, 397,
  726

\bibitem[\protect\citeauthoryear{{Liu}, {Xue}, {Fang}, {van de Ven}, {Wu},
  {Smith} \& {Carrell}}{{Liu} et~al.}{2012}]{2012ApJ...753L..24L}
{Liu} C.,  {Xue} X.,  {Fang} M.,  {van de Ven} G.,  {Wu} Y.,  {Smith} M.~C.,
  {Carrell} K.,  2012, \apjl, 753, L24


\bibitem[\protect\citeauthoryear{{Long} \& {Mao}}{{Long} \&
  {Mao}}{2010}]{2010MNRAS.405..301L}
{Long} R.~J.,  {Mao} S.,  2010, \mnras, 405, 301

\bibitem[\protect\citeauthoryear{{Long} \& {Mao}}{{Long} \&
  {Mao}}{2012}]{2012MNRAS.tmp.2408L}
{Long} R.~J.,  {Mao} S.,  2012, \mnras, p.~2408

\bibitem[\protect\citeauthoryear{{Long}, {Mao}, {Shen} \& {Wang}}{{Long}
  et~al.}{2013}]{2013MNRAS.428.3478L}
{Long} R.~J.,  {Mao} S.,  {Shen} J.,    {Wang} Y.,  2013, \mnras, 428, 3478

\bibitem[\protect\citeauthoryear{{Mao} \& {Paczy{\'n}ski}}{{Mao} \&
  {Paczy{\'n}ski}}{2002}]{2002MNRAS.337..895M}
{Mao} S.,  {Paczy{\'n}ski} B.,  2002, \mnras, 337, 895

\bibitem[\protect\citeauthoryear{{Martinez-Valpuesta} \&
  {Gerhard}}{{Martinez-Valpuesta} \& {Gerhard}}{2011}]{2011ApJ...734L..20M}
{Martinez-Valpuesta} I.,  {Gerhard} O.,  2011, \apjl, 734, L20

\bibitem[\protect\citeauthoryear{{Merritt} \& {Fridman}}{{Merritt} \&
  {Fridman}}{1996}]{1996ApJ...460..136M}
{Merritt} D.,  {Fridman} T.,  1996, \apj, 460, 136

\bibitem[\protect\citeauthoryear{{Minchev}, {Boily}, {Siebert} \&
  {Bienayme}}{{Minchev} et~al.}{2010}]{2010MNRAS.407.2122M}
{Minchev} I.,  {Boily} C.,  {Siebert} A.,    {Bienayme} O.,  2010, \mnras, 407,
  2122

\bibitem[\protect\citeauthoryear{{Minchev}, {Nordhaus} \& {Quillen}}{{Minchev}
  et~al.}{2007}]{2007ApJ...664L..31M}
{Minchev} I.,  {Nordhaus} J.,    {Quillen} A.~C.,  2007, \apjl, 664, L31

\bibitem[\protect\citeauthoryear{{Morganti} \& {Gerhard}}{{Morganti} \&
  {Gerhard}}{2012}]{2012MNRAS.tmp.2607M}
{Morganti} L.,  {Gerhard} O.,  2012, \mnras, p.~2607

\bibitem[\protect\citeauthoryear{{Paczynski}, {Stanek}, {Udalski}, {Szymanski},
  {Kaluzny}, {Kubiak}, {Mateo} \& {Krzeminski}}{{Paczynski}
  et~al.}{1994}]{1994ApJ...435L.113P}
{Paczynski} B.,  {Stanek} K.~Z.,  {Udalski} A.,  {Szymanski} M.,  {Kaluzny} J.,
   {Kubiak} M.,  {Mateo} M.,    {Krzeminski} W.,  1994, \apjl, 435, L113

\bibitem[\protect\citeauthoryear{{Pfenniger}}{{Pfenniger}}{1984}]{1984A&A...14%
1..171P}
{Pfenniger} D.,  1984, \aap, 141, 171

\bibitem[\protect\citeauthoryear{{Quinn}, {Perrine}, {Richardson} \&
  {Barnes}}{{Quinn} et~al.}{2010}]{2010AJ....139..803Q}
{Quinn} T.,  {Perrine} R.~P.,  {Richardson} D.~C.,    {Barnes} R.,  2010, \aj,
  139, 803

\bibitem[\protect\citeauthoryear{{Rattenbury}, {Mao}, {Sumi} \&
  {Smith}}{{Rattenbury} et~al.}{2007}]{2007MNRAS.378.1064R}
{Rattenbury} N.~J.,  {Mao} S.,  {Sumi} T.,    {Smith} M.~C.,  2007, \mnras,
  378, 1064

\bibitem[\protect\citeauthoryear{{Reid}, {Menten}, {Zheng}, {Brunthaler},
  {Moscadelli}, {Xu}, {Zhang}, {Sato}, {Honma}, {Hirota}, {Hachisuka}, {Choi},
  {Moellenbrock} \& {Bartkiewicz}}{{Reid} et~al.}{2009}]{2009ApJ...700..137R}
{Reid} M.~J.,  {Menten} K.~M.,  {Zheng} X.~W.,  {Brunthaler} A.,  {Moscadelli}
  L.,  {Xu} Y.,  {Zhang} B.,  {Sato} M.,  {Honma} M.,  {Hirota} T.,
  {Hachisuka} K.,  {Choi} Y.~K.,  {Moellenbrock} G.~A.,    {Bartkiewicz} A.,
  2009, \apj, 700, 137

\bibitem[\protect\citeauthoryear{{Rich}, {Reitzel}, {Howard} \& {Zhao}}{{Rich}
  et~al.}{2007}]{2007ApJ...658L..29R}
{Rich} R.~M.,  {Reitzel} D.~B.,  {Howard} C.~D.,    {Zhao} H.,  2007, \apjl,
  658, L29

\bibitem[\protect\citeauthoryear{{Saha} \& {Naab}}{{Saha} \&
  {Naab}}{2013}]{2013arXiv1304.1667S}
{Saha} K.,  {Naab} T.,  2013, ArXiv e-prints

\bibitem[\protect\citeauthoryear{{Sambhus} \& {Sridhar}}{{Sambhus} \&
  {Sridhar}}{2002}]{2002A&A...388..766S}
{Sambhus} N.,  {Sridhar} S.,  2002, \aap, 388, 766

\bibitem[\protect\citeauthoryear{{Schwarzschild}}{{Schwarzschild}}{1979}]{1979%
ApJ...232..236S}
{Schwarzschild} M.,  1979, \apj, 232, 236

\bibitem[\protect\citeauthoryear{{Sevenster}, {Saha}, {Valls-Gabaud} \&
  {Fux}}{{Sevenster} et~al.}{1999}]{1999MNRAS.307..584S}
{Sevenster} M.,  {Saha} P.,  {Valls-Gabaud} D.,    {Fux} R.,  1999, \mnras,
  307, 584

\bibitem[\protect\citeauthoryear{{Shen} \& {Gebhardt}}{{Shen} \&
  {Gebhardt}}{2010}]{2010ApJ...711..484S}
{Shen} J.,  {Gebhardt} K.,  2010, \apj, 711, 484

\bibitem[\protect\citeauthoryear{{Shen}, {Rich}, {Kormendy}, {Howard}, {De
  Propris} \& {Kunder}}{{Shen} et~al.}{2010}]{2010ApJ...720L..72S}
{Shen} J.,  {Rich} R.~M.,  {Kormendy} J.,  {Howard} C.~D.,  {De Propris} R.,
  {Kunder} A.,  2010, \apjl, 720, L72

\bibitem[\protect\citeauthoryear{{Skokos}, {Patsis} \& {Athanassoula}}{{Skokos}
  et~al.}{2002}]{2002MNRAS.333..861S}
{Skokos} C.,  {Patsis} P.~A.,    {Athanassoula} E.,  2002, \mnras, 333, 861

\bibitem[\protect\citeauthoryear{{Sofue}, {Honma} \& {Omodaka}}{{Sofue}
  et~al.}{2009}]{2009PASJ...61..227S}
{Sofue} Y.,  {Honma} M.,    {Omodaka} T.,  2009, \pasj, 61, 227

\bibitem[\protect\citeauthoryear{{Soto}, {Kuijken} \& {Rich}}{{Soto}
  et~al.}{2012}]{2012A&A...540A..48S}
{Soto} M.,  {Kuijken} K.,    {Rich} R.~M.,  2012, \aap, 540, A48

\bibitem[\protect\citeauthoryear{{Soto}, {Rich} \& {Kuijken}}{{Soto}
  et~al.}{2007}]{2007ApJ...665L..31S}
{Soto} M.,  {Rich} R.~M.,    {Kuijken} K.,  2007, \apjl, 665, L31

\bibitem[\protect\citeauthoryear{{Spaenhauer}, {Jones} \&
  {Whitford}}{{Spaenhauer} et~al.}{1992}]{1992AJ....103..297S}
{Spaenhauer} A.,  {Jones} B.~F.,    {Whitford} A.~E.,  1992, \aj, 103, 297

\bibitem[\protect\citeauthoryear{{Springel}}{{Springel}}{2005}]{2005MNRAS.364.%
1105S}
{Springel} V.,  2005, \mnras, 364, 1105

\bibitem[\protect\citeauthoryear{{Stanek}, {Mateo}, {Udalski}, {Szymanski},
  {Kaluzny} \& {Kubiak}}{{Stanek} et~al.}{1994}]{1994ApJ...429L..73S}
{Stanek} K.~Z.,  {Mateo} M.,  {Udalski} A.,  {Szymanski} M.,  {Kaluzny} J.,
  {Kubiak} M.,  1994, \apjl, 429, L73

\bibitem[\protect\citeauthoryear{{Sumi}, {Wu}, {Udalski}, {Szyma{\'n}ski},
  {Kubiak}, {Pietrzy{\'n}ski}, {Soszy{\'n}ski}, {Wo{\'z}niak},
  {{\.Z}ebru{\'n}}, {Szewczyk} \& {Wyrzykowski}}{{Sumi}
  et~al.}{2004}]{2004MNRAS.348.1439S}
{Sumi} T.,  {Wu} X.,  {Udalski} A.,  {Szyma{\'n}ski} M.,  {Kubiak} M.,
  {Pietrzy{\'n}ski} G.,  {Soszy{\'n}ski} I.,  {Wo{\'z}niak} P.,
  {{\.Z}ebru{\'n}} K.,  {Szewczyk} O.,    {Wyrzykowski} {\L}.,  2004, \mnras,
  348, 1439

\bibitem[\protect\citeauthoryear{{Syer} \& {Tremaine}}{{Syer} \&
  {Tremaine}}{1996}]{1996MNRAS.282..223S}
{Syer} D.,  {Tremaine} S.,  1996, \mnras, 282, 223

\bibitem[\protect\citeauthoryear{{Udalski}, {Zebrun}, {Szymanski}, {Kubiak},
  {Pietrzynski}, {Soszynski} \& {Wozniak}}{{Udalski}
  et~al.}{2000}]{2000AcA....50....1U}
{Udalski} A.,  {Zebrun} K.,  {Szymanski} M.,  {Kubiak} M.,  {Pietrzynski} G.,
  {Soszynski} I.,    {Wozniak} P.,  2000, \actaa, 50, 1

\bibitem[\protect\citeauthoryear{{van den Bosch}, {van de Ven}, {Verolme},
  {Cappellari} \& {de Zeeuw}}{{van den Bosch}
  et~al.}{2008}]{2008MNRAS.385..647V}
{van den Bosch} R.~C.~E.,  {van de Ven} G.,  {Verolme} E.~K.,  {Cappellari} M.,
     {de Zeeuw} P.~T.,  2008, \mnras, 385, 647

\bibitem[\protect\citeauthoryear{{Vieira}, {Casetti-Dinescu}, {M{\'e}ndez},
  {Rich}, {Girard}, {Korchagin}, {van Altena}, {Majewski} \& {van den
  Bergh}}{{Vieira} et~al.}{2007}]{2007AJ....134.1432V}
{Vieira} K.,  {Casetti-Dinescu} D.~I.,  {M{\'e}ndez} R.~A.,  {Rich} R.~M.,
  {Girard} T.~M.,  {Korchagin} V.~I.,  {van Altena} W.,  {Majewski} S.~R.,
  {van den Bergh} S.,  2007, \aj, 134, 1432

\bibitem[\protect\citeauthoryear{{Vieira}, {Casetti-Dinescu}, {M{\'e}ndez},
  {Rich}, {Girard}, {Korchagin}, {van Altena}, {Majeski} \& {van den
  Bergh}}{{Vieira} et~al.}{2009}]{2009RMxAC..35..123V}
{Vieira} K.,  {Casetti-Dinescu} D.~I.,  {M{\'e}ndez} R.~A.,  {Rich} R.~M.,
  {Girard} T.~M.,  {Korchagin} V.~I.,  {van Altena} W.~F.,  {Majeski} S.~R.,
  {van den Bergh} S.,  2009, in Revista Mexicana de Astronomia y Astrofisica
  Conference Series Vol.~35 of Revista Mexicana de Astronomia y Astrofisica
  Conference Series, {Proper Motions in the Galactic Bulge: Plaut's Window}.
pp 123--124

\bibitem[\protect\citeauthoryear{{Widrow}, {Pym} \& {Dubinski}}{{Widrow}
  et~al.}{2008}]{2008ApJ...679.1239W}
{Widrow} L.~M.,  {Pym} B.,    {Dubinski} J.,  2008, \apj, 679, 1239

\bibitem[\protect\citeauthoryear{{Xue}, {Rix}, {Zhao}, {Re Fiorentin}, {Naab},
  {Steinmetz}, {van den Bosch}, {Beers}, {Lee}, {Bell}, {Rockosi}, {Yanny},
  {Newberg}, {Wilhelm}, {Kang}, {Smith} \& {Schneider}}{{Xue}
  et~al.}{2008}]{2008ApJ...684.1143X}
{Xue} X.~X.,  {Rix} H.~W.,  {Zhao} G.,  {Re Fiorentin} P.,  {Naab} T.,
  {Steinmetz} M.,  {van den Bosch} F.~C.,  {Beers} T.~C.,  {Lee} Y.~S.,  {Bell}
  E.~F.,  {Rockosi} C.,  {Yanny} B.,  {Newberg} H.,  {Wilhelm} R.,  {Kang} X.,
  {Smith} M.~C.,    {Schneider} D.~P.,  2008, \apj, 684, 1143

\bibitem[\protect\citeauthoryear{{Wang}, {Zhao}, {Mao} \& {Rich}}{{Wang}
  et~al.}{2012}]{2012MNRAS.427.1429W}
{Wang} Y.,  {Zhao} H.,  {Mao} S.,    {Rich} R.~M.,  2012, \mnras, 427, 1429

\bibitem[\protect\citeauthoryear{{Zhao}}{{Zhao}}{1994}]{1994PhDT.........5Z}
{Zhao} H.,  1994, PhD thesis, Columbia Univ.

\bibitem[\protect\citeauthoryear{{Zhao}, {Rich} \& {Biello}}{{Zhao}
  et~al.}{1996}]{1996ApJ...470..506Z}
{Zhao} H.,  {Rich} R.~M.,    {Biello} J.,  1996, \apj, 470, 506

\bibitem[\protect\citeauthoryear{{Zhao}, {Rich} \& {Spergel}}{{Zhao}
  et~al.}{1996}]{1996MNRAS.282..175Z}
{Zhao} H.,  {Rich} R.~M.,    {Spergel} D.~N.,  1996, \mnras, 282, 175

\bibitem[\protect\citeauthoryear{{Zhao}, {Spergel} \& {Rich}}{{Zhao}
  et~al.}{1995}]{1995ApJ...440L..13Z}
{Zhao} H.,  {Spergel} D.~N.,    {Rich} R.~M.,  1995, \apjl, 440, L13

\bibitem[\protect\citeauthoryear{{Zhao}}{{Zhao}}{1996}]{1996MNRAS.283..149Z}
{Zhao} H.~S.,  1996, \mnras, 283, 149

\end{thebibliography}
\label{lastpage}

\end{document}